\documentclass[aps,pra,twocolumn,groupedaddress,showpacs]{revtex4-1}

\usepackage{graphicx}
\usepackage{longtable}

\begin{document}


\title{Multiqubit entanglement of a general input state}


\author{E.C. Behrman}
\email[]{elizabeth.behrman@wichita.edu}
\affiliation{Department of Mathematics and Physics, Wichita State University, Wichita, KS 67260-0033}
\author{J.E. Steck}
\email[]{james.steck@wichita.edu}
\affiliation{Department of Aerospace Engineering, Wichita State University, Wichita, KS 67260-0044}


\date{\today}

\begin{abstract}
Measurement of entanglement remains an important problem for quantum information. We present the design and simulation of an experimental  method for entanglement estimation for a general multiqubit state. The system can be in a pure or a mixed state, and it need not be ``close'' to any particular state. Our method, based on dynamic learning, does not require prior state reconstruction or lengthy optimization. Results for three-qubit systems compare favorably with known entanglement measures. The method is then extended to four- and five-qubit systems, with relative ease. As the size of the system grows the amount of training necessary diminishes, raising hopes for applicability to large computational systems.
\end{abstract}


\maketitle

\section{Introduction\label{intro}}

Because entanglement is crucial to most quantum information applications, convenient and accurate methods for its measurement and calculation are essential.  Alas, though considerable work by a large number of very intelligent people has made considerable progress (see, e.g.,\cite{bennett2,wootters,ghz,wootters02,vedral,tamaryan,park,lohmayer,filip,coffman,jung, bai}), we have as yet no general method, of universal applicability, much less one which is easy to implement.

For 2-qubit systems the problem is fairly well understood, from more than one point of view. For example, the ``entanglement of formation''\cite{bennett2,wootters} approach is based on the idea of construction: how much entanglement is necessary in order to reconstruct this particular state? This gives us a general measure as an analytic formula, so that, if one knows the density matrix, one can easily calculate the entanglement. However, generalizations to larger systems are difficult\cite{ghz}. Various geometric methods\cite{vedral, tamaryan} use an extremely appealing idea of ``closeness'' to the subspace of separable states: one constructs some distance measure in the state space, then, performs a minimization calculation to find the distance of closest approach. Product states, of course, have zero distance to that space, and have therefore zero entanglement. Many entanglement witnesses\cite{park} are based on this same idea. The drawback is, of course, the optimization calculation, which rapidly becomes impractically difficult for large systems. In addition, these approaches assume that one knows the density matrix. One might not. And reconstruction of the density matrix, like optimization, is not easy to generalize to large systems. Most witnesses also require that the state of the system be ``close'', in some sense, to a given, known state\cite{filip}.

Clearly it would be of great interest to have a general, experimental method for estimation of the entanglement, for which one would not have to know or to reconstruct the density matrix, which would not require lengthy optimization procedures, and which could be applied to any state. We present here such a method. We demonstrate that our method works for large classes of states of 2-qubit, 3-qubit, 4-qubit, and 5-qubit systems, and show that extension to larger systems is clearly achievable.

\section{Dynamic learning: quantum neural network (QNN)\label{qnn}}

We begin with the Schr\"{o}dinger equation:
\begin{equation}
\frac{d \rho}{dt} = \frac{1}{i \hbar}[H, \rho] 
\label{schr}
\end{equation}
where $\rho$ is the density matrix and $H$ is the Hamiltonian, whose formal solution\cite{peres} is 
\begin{equation}
\rho(t) = \exp (iLt) \rho (0),
\label{schrsoln}
\end{equation}
where $L$ is the Liouville operator. We consider an N-qubit system whose Hamiltonian is:
\begin{equation}
H   = \sum_{\alpha=1}^{N}  K_{\alpha} \sigma_{x\alpha} + \varepsilon_{\alpha} \sigma_{z\alpha} + \sum_{\alpha\neq\beta=1}^{N}\zeta_{\alpha\beta} \sigma_{z\alpha} \sigma_{z\beta}
\label{ham}
\end{equation}
where $\{ \sigma \}$ are the Pauli operators corresponding to each of the qubits, $\{K \}$ are the tunneling amplitudes, $\{ \varepsilon \}$  are the biases, and $\{ \zeta \}$, the qubit-qubit couplings. We choose the usual ``charge basis '', in which each qubit's state is given as 0 or 1; for a system of N qubits there are $2^{N}$ states, each labelled by a bit string each of whose numbers corresponds to the state of each qubit, in order. The amplitude for each qubit to tunnel to its opposing state (i.e., switch between the 0 and 1 states) is its $K$ value; each qubit has an external bias represented by its $\varepsilon$ value; and each qubit is coupled to each of the other qubits, with a strength represented by the appropriate $\zeta$ value. Note that, for example, the operator $\sigma_{xA} = \sigma_{x}\otimes I \: ... \: \otimes I$, where there are (N-1) outer products, acts nontrivially only on qubit A. 

The parameters $\{ K,\varepsilon,\zeta \}$ direct the time evolution of the system in the sense that, if one or more of them is changed, the way a given state will evolve in time will also change, because of Eqs.~1-3. This is the basis for using our quantum system as a neural network. There is a mathematical isomorphism between Eq.~\ref{schrsoln} and the equation for information propagation in a neural network: $\phi_{output}= F_{W} \phi_{input }$, where $\phi_{output}$ is the output vector of the network, $\phi_{input}$ the input vector, and $F_{W}$ the network operator, which depends on the neuron connectivity weight matrix $W$. Here the role of the input vector is the initial density matrix $\rho(0)$, the role of the output is the density matrix at the final time, $\rho(t_{f})$, and the role of the ``weights'' of the network is played by the parameters of the Hamiltonian, $\{ K,\varepsilon,\zeta  \}$, all of which can be adjusted experimentally as functions of time\cite{yamamoto}. By adjusting the parameters using a neural network type learning algorithm we can train the system to evolve in time to a set of particular final states at the final time $t_{f}$, in response to a corresponding (one-to-one) set of given inputs. Because the time evolution is quantum mechanical (and, we assume, coherent), a quantum mechanical function, like an entanglement witness of the initial state, can be mapped to an observable of the system's final state, a measurement made at the final time $t_{f}$. Complete details, including a derivation of the quantum dynamic learning paradigm using backpropagation\cite{lecun} in time\cite{werbos}, are given in \cite{behrmanqic}.

The time evolution of the quantum system is calculated by integrating the Schr\"{o}dinger equation numerically in MATLAB 
Simulink, using ODE4 (Runge-Kutta), with a fixed integration step size of 0.05 ns\cite{matlab}. The system was initialized (prepared in) each input state, in turn, then allowed to evolve for 300 ns. A measurement is then made at the final time. All of the parameters $\{ K,\varepsilon,\zeta \}$ were taken to be functions of time; this was done in the simulation by allowing them to change to a different constant value every 75 ns (i.e., four ``time chunks''.)  Discretization error for the numerical integration was checked by redoing the calculations with a timestep of a tenth the size; results were not affected.  For the backpropagation learning, the output error needs to be back-propagated through time\cite{werbos}, so the integration has to be carried out from $t_{f}$ to 0.  To implement this in MATLAB Simulink, a change of variable is made by letting $t' = t_{f}-t$, and running this simulation forward in $t'$ in Simulink.

\section{Why this is not quantum control\label{nqc}}
More generally, a nonlinear system can be written as
\begin{eqnarray}
 {\dot{x}=f(x,u,p)} \\ {y=g(x,u)} 
 \label{contr1}
 \end{eqnarray}
where x is the state, f is a nonlinear function, g is an output measure, u is the system external control input, and y is the system output. x, u, and y are all functions of time.  Notice from Eq.~\ref{schr} that our quantum system is of the form $\dot{x} = Ax$, and the (complex) operator $A$ contains the parameters p = $\{ K,\varepsilon,\zeta \}$. In general p can be a set of a priori \underline{fixed} system parameters, (e.g. the mass, spring constant, and damping coefficient, in a mass spring damper system.) u is a control input that is not fixed ahead of time, and can be calculated on the fly and varied as the system is running, i.e., for every different run of the system, u can be recalculated based, usually, on the current state of the system x(t) as it is going from x(0) to x($t_{f}$).  So u can be different for different initial conditions (x(0) values). While the set of parameters p can be functions of time as the system runs, they are fixed at the same values every time the system is run going from any x(0) to x($t_{f}$), once determined via the network learning process.

A control problem, then, would be to calculate u on the fly over a period of time in order to make the system states x or output y behave in a certain way.  An optimal control problem would be to choose u to minimize some cost function J(x,u) or J(y,u). Given initial conditions x(0) and u, the system can be integrated from t=0 to t=$t_{f}$ so that
\begin{eqnarray}
x(t_{f})=F(x(0),u,p) \\ y(t_{f})=g(x(t_{f}),u(t_{f}))
  \end{eqnarray}
Our quantum system, on the other hand, has no external input, so it is written
\begin{eqnarray} 
x(t_{f})=Q(x(0),p) \\ y(t_{f})=m(x(t_{f})) 
\label{qcont}
\end{eqnarray}
where now Q defines the quantum system and m is some output measurement made on the system at the final time.  The quantum computation problem is to find values for the specific fixed parameters p such that the quantum system approximates a given function $\hat{s}=h(\hat{r})$at a finite number of points, m, in the domain, that is, $y(t_{f})=\hat{s}_{i} \mathop{}\limits_{}$ when $\mathop{}\limits_{} x(0)=\hat{r}_{i} \mathop{}\limits_{} i=1,m$.

This is not a control problem, but a problem of finding a finite dimensional (with dimension equal to the number of parameters in p) subspace approximation of the function h. Interpreted as a quantum neural network, the weights p are found through a learning algorithm, where $[\hat{r}_{i} ,\hat{s}_{i} ]$ i=1,m are the training pairs in a training set, and the cost function to be minimized is $\hat{J}=[\hat{s}_{i} -y_{i} (t_{f})]^{2} $ for each and every i=1,m, where from Eqs.~8-9, $y{}_{i}(t_{f}$)=m(Q($x{}_{i}$(0),p)) is the output for the system started in initial condition $x{}_{i}$(0)=r${}_{i}$.

\section{Entanglement\label{ent}}

Almost all the information contained within a system of 3 qubits is contained within the 2-particle reduced density matrices\cite{wootters02}. Thus it makes sense to train our entanglement measure, to begin with, to the set of pairwise entanglements. Fortunately in previous work\cite{behrmanqic} we have already found a set of parameters that successfully maps the input state of a two-qubit system to a good approximation of the entanglement of formation, via the qubit-qubit correlation function at the final time, $\langle \sigma_{zA}(t_{f})\sigma_{zB}(t_{f})\rangle^{2}$. The parameters were found by training with a set of just four inputs, as shown in Table~\ref{inputs}: a fully entangled state, a ``flat'' state (equal amounts of all basis states), a correlated product state, and a partially entangled state. 

For the three-qubit state ABC there are three possible pairs: AB, AC, and BC; and therefore three different pairwise entanglements to train. We therefore used\cite{behrmannabic} a set of twelve training pairs: the four in Table~\ref{inputs} for each pair, outerproducted with $|0\rangle$ for the unpaired qubit. So, for example, the $Bell_{AC}$ state was taken to be $\frac{1}{\sqrt{2}}(|000\rangle+|101\rangle)$, and was trained to give an output of 1 for the network output measure $O_{AC}=\langle \sigma_{zA}(t_{f}) \sigma_{zC}(t_{f})\rangle^{2}$ and zero for each of the output measures $O_{AB}=\langle \sigma_{zA}(t_{f})\sigma_{zB}(t_{f})\rangle^{2}$ and $O_{BC}=\langle \sigma_{zB}(t_{f}) \sigma_{zC}(t_{f})\rangle^{2}$.  Table~\ref{train3} shows the target outputs and average calculated outputs after training; detailed procedures and results are in \cite{behrmannabic}. Note that we report here the \underline{averages}; e.g., the $Bell_{\alpha\beta}$ column numbers are the averages over $\alpha\beta =$ AB, AC, and BC. (``Non-$\alpha\beta$'' refers to the other two outputs for each pair; there are six of these numbers.)  

\begin{table}[!t]
\caption{Input matrices for training the pairwise quantum neural network (QNN) entanglement witnesses. Each column is an input state, showing relative amounts of each of the basis states. The N-qubit system was trained on this set of four for each of the $\left(\begin{array}{c} N \\ 2 \end{array}\right)$ pairs; this constituted the ``pairwise'' training. \label{inputs}}                                                                                                                  
\begin{ruledtabular}
\begin{tabular}{|l| c c c c|} 
Input       & Bell  &   Flat   & Corr.  &  P \\
\hline
$ |00\rangle $  & 1 & 1 & 0   & 1   \\
$ |01\rangle $  & 0 & 1 & 0   & 1  \\
$ |10\rangle $  & 0 & 1 & 0.5 & 1  \\
$ |11\rangle $  & 1 & 1 & 1   & 0  \\  
\end{tabular} 
\end{ruledtabular}
\end{table}

\begin{table*}[!t]
\caption{Stages 2 \& 3: Three-qubit QNN entanglement targets and calculated average outputs. Each column is an input state corresponding to the respective column in Table~\ref{inputs}; each row corresponds to a target function for the specified entanglement function. Stage 2 started from the trained 2-qubit system's parameters (stage 1)\cite{behrmanqic}, and trained the three-qubit system to the set of pairwise entanglement training pairs(12 training pairs, 3 outputs per pair); Stage 3 used the pairwise set plus the GHZ state (13 training pairs, 4 outputs per pair). The double vertical line separates the states included in the training set from those only tested on. Nonzero target-training pairs are boldfaced for easy comparison. RMS errors were: $1.2\times 10^{-3}$ for stage 2, and  $1.8\times 10^{-3}$ for stage 3.\label{train3}} 
\begin{ruledtabular} 
\begin{tabular}{|l|l | l l l l ||l|}
& $O$ & $Bell_{\alpha\beta}$  &  Flat  & Corr.  &  P & $GHZ_{3}$ \\ \hline
 Targets: & $\alpha\beta$ & \bf{1} & 0 & 0 & \bf{0.44}   & -  \\
 Stage 2 & non-$\alpha\beta$           & 0 & 0 & 0 & 0      & - \\ \hline
 Trained & $\alpha\beta$ & \bf{0.9939} & 0.0001 & 0.0004 & \bf{0.4392} & -  \\
         & non-$\alpha\beta$        & 0.0006 & 0.0006 & 0.0014 & 0.0006  & 0.0032 \\
              & ABC        & 0.0081 & 0.0001 & 0.0001 & 0.0037 & 0.5735 \\   
\end{tabular}
\begin{tabular}{|l|l | l l l l |l|} 
 Targets: & $\alpha\beta$ & \bf{1} & 0 & 0 & \bf{0.44}   & -   \\
 Stage 3 & non-$\alpha\beta$           & 0 & 0 & 0 & 0   & 0  \\
      & ABC               & 0 & 0 & 0 & 0      & \bf{1}  \\ \hline 
 Trained & $\alpha\beta$  & \bf{0.9948} & 0.0001 & 0.0009 & \bf{0.4389}  & -  \\
             & non-$\alpha\beta$     & 0.0005 & 0.0012 & 0.0004 & 0.0022 & 0.0035 \\
             & ABC        & 0.0013 & 0.0006 & 0.0001 & 0.0012 & \bf{0.9883} \\ 
\end{tabular} 
\end{ruledtabular}
\end{table*}

After the pairwise training was accomplished for the 3-qubit system, we then trained the residual\cite{coffman} entanglement, as well. We did this by expanding the training set: adding one additional training pair, the three-way entangled Greenberger-Horne-Zeilinger (GHZ)\cite{ghz} state for three qubits, $\frac{1}{\sqrt{2}}(|000\rangle+|111\rangle)$, for which we specified targets of zero for all three pairwise correlation functions, and 1 for the three-point correlation function $O_{ABC}=\langle\sigma_{zA}(t_{f})\sigma_{zB}(t_{f})\sigma_{zC}(t_{f})\rangle^{2}$. The bottom row in Table~\ref{train3} displays these targets and calculated average outputs. 

Once trained, the parameters that are found can be used to evaluate the entanglement of any desired state. (This will tell us whether we actually have an estimate of the entanglement, or have just been curvefitting the proverbial elephant.) Testing was therefore done on a large number of states not represented in the training set, including fully entangled states, partially entangled states, product (unentangled) states, and also mixed states. Detailed results were presented in \cite{behrmannabic};  while errors were significantly larger, it was clear from the output matrix that there is definite separation, in two senses: first, that it is easy to see where the pairwise entanglement is (e.g., to distinguish between a state with AB entanglement and one with BC entanglement); and, second, that it is easy to differentiate among unentangled, partially entangled, and fully entangled states. In other words, while this is not, strictly speaking, an entanglement measure, it is quite a good witness.

It was natural to take this one step further, to a system of four qubits. Again, we started from the trained parameters for the smaller (3-qubit) system, and simply copied the new sets necessary (one tunneling parameter, one bias parameter, and three coupling parameters) from the trained sets already found. Results are shown in Tables~\ref{train44}, ~\ref{train45}, and ~\ref{train46}. Again, we worked in stages: starting with a training set of 24 for the pairwise training (6 pairwise Bell states, 6 pairwise flat states, 6 pairwise correlated unentangled states, and 6 pairwise partly entangled). Once those were trained, we added four more pairs corresponding to the four distinct 3-way GHZ states; after those were trained, we added also the single 4-way GHZ state. Testing was then done, as with the 3-qubit system, on a similar (but expanded) set of states. Errors were slightly larger but the results maintained good separation. Buoyed by confidence in our method we then successfully expanded to a 5-qubit system, using an exactly similar procedure. Training results are shown in Tables~\ref{train578}, \ref{train59} and ~\ref{train510}. Final values for the parameters are in Table~\ref{param}. 

\begin{table*}[!t]
\caption{Stage 4: Four-qubit QNN entanglement targets and calculated average outputs. Stage 4 trained the four-qubit system to  the set of 24 pairwise entanglement training pairs. RMS error = $2.2\times 10^{-3}$ (24 training pairs, 6 outputs per pair). \label{train44}}  
\begin{ruledtabular}
\begin{tabular}{|l|l|  l l l l ||l|l|} 
 & $O$ & $Bell_{\alpha\beta}$  &  Flat  & Corr.  &  P & $GHZ_{\alpha\beta\gamma}$ & $GHZ_{4}$\\ \hline
 Targets: & $\alpha\beta$ & \bf{1} & 0 & 0 & \bf{0.44} & - &  -  \\
 Stage 4 & non-$\alpha\beta$& 0 & 0 & 0 & 0    & - &  -  \\ \hline

 Trained & $\alpha\beta$  & \bf{0.9830} & 0.0002 & 0.0008 & \bf{0.4426}  & -  & - \\
   & non-$\alpha\beta$    & 0.0009 & 0.0024 & 0.0028 & 0.0017 & 0.0023 & 0.0032 \\
  & $\alpha\beta\gamma$   & -      & -      & -      & -      & 0.8566 & - \\
& non-$\alpha\beta\gamma$ & 0.0007 & 0.0014 & 0.0012 & 0.0003 & 0.0001 & 0.0004 \\
& ABCD                    & 0.0021 & 0.0010 & 0.0000 & 0.0037 & 0.0020 & 0.9089 \\ 
\end{tabular}
\end{ruledtabular}
\end{table*}

\begin{table*}[!t]
\caption{Stage 5: four-qubit QNN entanglement targets and calculated average outputs. Stage 5 added the four 3-way GHZ states to the training set of Table~\ref{train44}. RMS error = $1.9\times 10^{-3}$ (28 training pairs, 10 outputs per pair). \label{train45}}  
\begin{ruledtabular}
\begin{tabular}{|l|l|  l l l l |l||l|} 
 & $O$ & $Bell_{\alpha\beta}$  &  Flat  & Corr.  &  P & $GHZ_{\alpha\beta\gamma}$ & $GHZ_{4}$\\ \hline
 Targets: & $\alpha\beta$ & \bf{1} & 0 & 0 & \bf{0.44} & - & -  \\
 Stage 5 & non-$\alpha\beta$ & 0 & 0 & 0 & 0    & 0 & -  \\
  & $\alpha\beta\gamma$   & - & - & - & -    & \bf{1} & -  \\
& non-$\alpha\beta\gamma$ & 0 & 0 & 0 & 0    & 0 & -  \\ \hline 
 Trained & $\alpha\beta$  & \bf{0.9731} & 0.0001 & 0.0005 & \bf{0.4360}  & -  & - \\
   & non-$\alpha\beta$    & 0.0006 & 0.0023 & 0.0024 & 0.0019 & 0.0023 & 0.0038 \\
 & $\alpha\beta\gamma$   & - & - & -& - & \bf{0.9728} & - \\
 & non-$\alpha\beta\gamma$ & 0.0003 & 0.0014 & 0.0012 & 0.0003 & 0.0001 & 0.0007 \\
& ABCD                    & 0.0014 & 0.0012 & 0.0000 & 0.0036 & 0.0010 & 0.7649 \\ 
\end{tabular}
\end{ruledtabular}
\end{table*}

\begin{table*}[!t]
\caption{Stage 6: four-qubit QNN entanglement targets and calculated average outputs. Stage 6 added the one 4-way GHZ state to the training set of stage 5. RMS error = $2.2\times 10^{-3}$ (29 training pairs, 11 outputs per pair). \label{train46}} 
\begin{ruledtabular} 
\begin{tabular}{|l|l|  l l l l |l|l|} 
 & $O$ & $Bell_{\alpha\beta}$  &  Flat  & Corr.  &  P & $GHZ_{\alpha\beta\gamma}$ & $GHZ_{4}$\\ \hline
 Targets: & $\alpha\beta$  & \bf{1} & 0 & 0 & \bf{0.44} & - & -  \\
 Stage 6 & non-$\alpha\beta$ & 0 & 0 & 0 & 0    & 0 & 0  \\
    & $\alpha\beta\gamma$  & - & - & - & -    & \bf{1} & -  \\ 
& non-$\alpha\beta\gamma$  & 0 & 0 & 0 & 0    & 0 & 0 \\
    & ABCD                 & 0 & 0 & 0 & 0    & 0 & \bf{1}  \\ \hline  
 Trained & $\alpha\beta$  & \bf{0.9682} & 0.0000 & 0.0002 & \bf{0.4341}  & -  & - \\
 & non-$\alpha\beta$       & 0.0002 & 0.0009 & 0.0011 & 0.0007 & 0.0010 & 0.0017 \\
 & $\alpha\beta\gamma$    & - & - & - & - & \bf{0.9592} & - \\
 & non-$\alpha\beta\gamma$& 0.0002 & 0.0005 & 0.0004 & 0.0002 & 0.0001 & 0.0003 \\
 & ABCD                   & 0.0004 & 0.0003 & 0.0000 & 0.0011 & 0.0014 & \bf{0.9488} \\  
\end{tabular} 
\end{ruledtabular}
\end{table*}

\begin{table*}
\caption{Stages 7 \& 8: five-qubit QNN entanglement targets and calculated average outputs. Stage 7 trained the five-qubit system to the set of 40 pairwise entanglement states for the 5-qubit system; Stage 8 added the ten 3-way GHZ states. RMS errors were: $5.6\times 10^{-4}$ for stage 7 (40 training pairs, 10 outputs per pair), and $7.9\times 10^{-4}$ for stage 8 (50 training pairs, 20 outputs per pair). \label{train578}} 
\begin{ruledtabular} 
\begin{tabular}{|l|l|  l l l l ||l|l|l|} 
 &       $O$                 & $Bell_{\alpha\beta}$  &  Flat  & Corr.  &  P & $GHZ_{\alpha\beta\gamma}$ & $GHZ_{\alpha\beta\gamma\delta}$ & $GHZ_{5}$\\ \hline
 Targets:       & $\alpha\beta$ & \bf{1}      & 0      & 0      & \bf{0.44}   & -  &  - & - \\
 Stage 7 & non-$\alpha\beta$    & 0      & 0      & 0      & 0      & -  &  - & - \\ \hline

 Trained & $\alpha\beta$        & \bf{0.9908} & 0.0000 & 0.0003 & \bf{0.4400} & -  & -  & - \\
& non-$\alpha\beta$             & 0.0003 & 0.0005 & 0.0010 & 0.0003 & 0.0003 & 0.0005 & 0.0007 \\
& $\alpha\beta\gamma$           & -      & -      & -      & -      & 0.7055 & -  & - \\
& non-$\alpha\beta\gamma$       & 0.0002 & 0.0003 & 0.0003 & 0.0001 & 0.0001 & 0.0001 & 0.0022  \\
& $\alpha\beta\gamma\delta$     & -      & -      & -      & -      & - & 0.7806 &  - \\
& non-$\alpha\beta\gamma\delta$ & 0.0004 & 0.0001 & 0.0000 & 0.0003 & 0.0015 & 0.0001 & 0.0013 \\ 
& ABCDE                         & 0.0000 & 0.0000 & 0.0000 & 0.0000 & 0.0005 & 0.0039 & 0.8307 \\  
\end{tabular}
\begin{tabular}{|l|l|  l l l l |l||l|l|}
 Targets: & $\alpha\beta$       & \bf{1}      & 0      & 0      & \bf{0.44}   & -       & -  & - \\
 Stage 8 & non-$\alpha\beta$    & 0      & 0      & 0      & 0      & 0       & -  & - \\
& $\alpha\beta\gamma$           & -      & -      & -      & -      & \bf{1}       & -  & - \\
& non-$\alpha\beta\gamma$       & 0      & 0      & 0      & 0      & 0       & -  & - \\ \hline 
 Trained & $\alpha\beta$        & \bf{0.9754} & 0.0000 & 0.0002 & \bf{0.4323} & -       & -  & - \\
& non-$\alpha\beta$             & 0.0001 & 0.0005 & 0.0005 & 0.0005 & 0.0003  & 0.0003 & 0.0001 \\
& $\alpha\beta\gamma$           & -      & -      & -      & -      & \bf{0.9765}  & -  & - \\
& non-$\alpha\beta\gamma$       & 0.0001 & 0.0002 & 0.0001 & 0.0001 & 0.0001  & 0.0006 & 0.0006 \\
& $\alpha\beta\gamma\delta$     & -      & -      & -      & -      & -       & 0.7597 & - \\
& non-$\alpha\beta\gamma\delta$ & 0.0002 & 0.0000 & 0.0000 & 0.0002 & 0.0004  & 0.0002 & 0.0085 \\ 
& ABCDE                         & 0.0000 & 0.0000 & 0.0000 & 0.0000 & 0.0000  & 0.0090 & 0.0612 \\ 

\end{tabular} 
\end{ruledtabular}
\end{table*}
\begin{table*}
\caption{Stage 9: five qubit QNN entanglement targets and calculated average outputs. Stage 9 added the five 4-way GHZ states to the training set of stage 8. RMS error = $1.4\times 10^{-3}$ (55 training pairs, 25 outputs per pair).\label{train59}}  
\begin{ruledtabular}
\begin{tabular}{|l|l|  l l l l |l|l||l|} 
 & $O$ & $Bell_{\alpha\beta}$  &  Flat  & Corr.  &  P & $GHZ_{\alpha\beta\gamma}$ & $GHZ_{\alpha\beta\gamma\delta}$ & $GHZ_{5}$\\ \hline
 Targets: & $\alpha\beta$       & \bf{1}      & 0      & 0      & \bf{0.44}   & -       & -  & - \\
 Stage 9 & non-$\alpha\beta$    & 0      & 0      & 0      & 0      & 0       & 0  & - \\
    & $\alpha\beta\gamma$       & -      & -      &  -     & -      & \bf{1}       & -  & - \\ 
& non-$\alpha\beta\gamma$       & 0      & 0      & 0      & 0      & 0       & 0  & - \\
& $\alpha\beta\gamma\delta$     & -      & -      & -      & -      & -       & \bf{1}  & -  \\ 
& non-$\alpha\beta\gamma\delta$ & 0      & 0      & 0      & 0      & 0       & 0  & -  \\ \hline  
 Trained & $\alpha\beta$        & \bf{0.9647} & 0.0000 & 0.0001 & \bf{0.4315} & -       & -      & - \\
 & non-$\alpha\beta$            & 0.0000 & 0.0002 & 0.0002 & 0.0002 & 0.0001  & 0.0001 & 0.0002 \\
 & $\alpha\beta\gamma$          & -      & -      & -      & -      & \bf{0.9574} & -      & - \\
 & non-$\alpha\beta\gamma$      & 0.0000 & 0.0001 & 0.0000 & 0.0001 & 0.0000 & 0.0002 & 0.0005 \\
& $\alpha\beta\gamma\delta$     &-     & -      & -      & -      & -        & \bf{0.9630} & - \\
& non-$\alpha\beta\gamma\delta$ & 0.0001 & 0.0000 & 0.0000 & 0.0000 & 0.0002 & 0.0001 & 0.0015 \\ 
& ABCDE                         & 0.0000 & 0.0000 & 0.0000 & 0.0000 & 0.0002 & 0.0022 & 0.1855 \\ 
\end{tabular}
\end{ruledtabular}
\end{table*}

\begin{table*}
\caption{Stage 10: five-qubit QNN entanglement targets and calculated average outputs. Stage 10 added the one 5-way GHZ state to the training set of stage 9.  RMS error = $2.4\times 10^{-3}$ (56 training pairs, 26 outputs per pair).\label{train510}}  
\begin{ruledtabular}
\begin{tabular}{|l|l|  l l l l |l|l|l|}
 & $O$ & $Bell_{\alpha\beta}$  &  Flat  & Corr.  &  P & $GHZ_{\alpha\beta\gamma}$ & $GHZ_{\alpha\beta\gamma\delta}$ & $GHZ_{5}$\\ \hline
Targets: & $\alpha\beta$        & \bf{1} & 0 & 0 & \bf{0.44} & - & -  & - \\
Stage 10 & non-$\alpha\beta$    & 0 & 0 & 0 & 0    & 0 & 0  & 0 \\
    & $\alpha\beta\gamma$       & - & - & - & -    & \bf{1} & -  & - \\ 
& non-$\alpha\beta\gamma$       & 0 & 0 & 0 & 0    & 0 & 0  & 0 \\
& $\alpha\beta\gamma\delta$     & - & - & - & -    & - & \bf{1}  & -  \\ 
& non-$\alpha\beta\gamma\delta$ & 0 & 0 & 0 & 0    & 0 & 0  & 0  \\ 
& ABCDE                         & 0 & 0 & 0 & 0    & 0 & 0  & \bf{1}  \\ \hline  
 Trained & $\alpha\beta$        & \bf{0.9148} & 0.0000 & 0.0000 & \bf{0.4295} & -      & -      & -     \\
 & non-$\alpha\beta$            & 0.0000 & 0.0001 & 0.0001 & 0.0001 & 0.0001 & 0.0000 & 0.0003 \\
 & $\alpha\beta\gamma$          & -      & -      & -      & -      & \bf{0.9195} & -      & -      \\
 & non-$\alpha\beta\gamma$      & 0.0000 & 0.0005 & 0.0004 & 0.0002 & 0.0000 & 0.0000 & 0.0005 \\
& $\alpha\beta\gamma\delta$     & -      & -      & -      & -      & -      & \bf{0.9195} & -      \\
& non-$\alpha\beta\gamma\delta$ & 0.0000 & 0.0000 & 0.0000 & 0.0000 & 0.0001 & 0.0000 & 0.0004 \\ 
& ABCDE                         & 0.0000 & 0.0000 & 0.0000 & 0.0000 & 0.0001 & 0.0001 & \bf{0.8258} \\ 
\end{tabular} 
\end{ruledtabular}
\end{table*}

\begin{table}[!t]
\caption{Final (stage 10) trained parameters for entanglement of 5-qubit system, in MHz. Each row is a different parameter, as labelled. Each parameter is a function of time; the first row shows the timeslice number. \label{param} }
\begin{ruledtabular}
\begin{tabular}{|l| c c c c|} 
 t & 1 & 2 & 3 & 4 \\ \hline
$K_{A}$ & 2.1733 & 2.4021 & 2.4688 & 2.4139 \\
$K_{B}$ & 2.5842 & 2.1591 & 2.2702 & 2.4710 \\
$K_{C}$ & 2.3984 & 2.3252 & 2.3501 & 2.4294 \\
$K_{D}$ & 2.3893 & 2.3128 & 2.3451 & 2.4327 \\
$K_{E}$ & 2.3926 & 2.3144 & 2.3446 & 2.4274 \\
$\varepsilon_{A}$ &   0.7443 & -0.1558 & -1.1248 & -0.7230 \\
$\varepsilon_{B}$ &   0.9903 & -0.4483 & -1.0216 & -0.4294 \\
$\varepsilon_{C}$ &   1.1027 &  -0.4998 & -0.8678 & -0.4752 \\
$\varepsilon_{D}$ &  0.7783  & -0.2453  & -1.1015 & -0.6125 \\
$\varepsilon_{E}$ &  0.7703  & -0.2472  & -1.0941 & -0.6116 \\
$\zeta_{AB}$ &   -0.3576 & 0.2797 &  -0.5974 & 0.1521 \\
$\zeta_{AC}$ &  -0.1761 & 0.1800 & -0.5796 & -0.0604 \\
$\zeta_{AD}$ &  -0.1687 & 0.1802 & -0.5558 & -0.0471 \\
$\zeta_{AE}$ &  -0.1563 & 0.1758 & -0.5324 & -0.0402 \\
$\zeta_{BC}$ &  -0.2228 &  0.1806 & -0.5462 & 0.0340 \\
$\zeta_{BD}$ & -0.2138 & 0.1808 & -0.5227 & 0.0431 \\
$\zeta_{BE}$ & -0.2026 & 0.1785 & -0.5014 & 0.0484 \\
$\zeta_{CD}$ &  -0.3052 &   0.2375 &  -0.4795 & 0.1959   \\
$\zeta_{CE}$ &  -0.2965 &   0.2358 &  -0.4638 & 0.1801  \\
$\zeta_{DE}$ &  -0.2896 &   0.2265 &  -0.4465 & 0.1771  \\ 
 
\end{tabular}
\end{ruledtabular}
\end{table}

\section{Results and discussion\label{res}}

\begin{figure}
\includegraphics[height=2.5in]{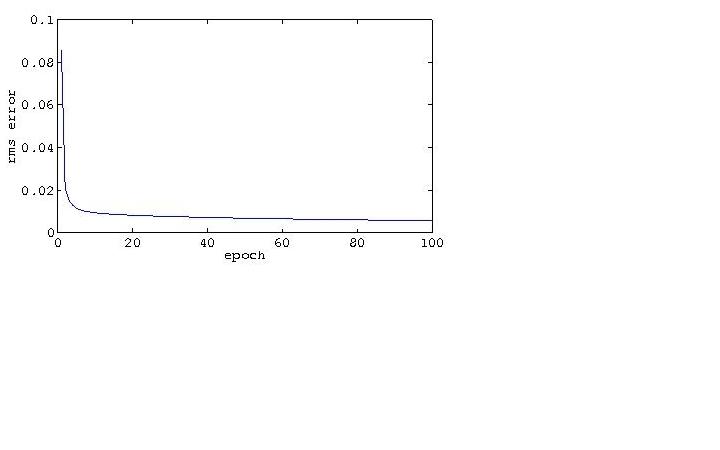} 
\caption{Root mean squared error per training pair as a function of epoch (pass through the training set), for the 5-qubit system on the pairwise training set (stage 7), summed over the 10 outputs per pair. \label{5q2err}}
\end{figure}

Training of the three-qubit system required considerable time: 5000 epochs (passes through the entire training set) to get pairwise entanglement from the trained results for the two-qubit system\cite{behrmannabic}. But training of the four-qubit system was (unexpectedly) easy. The training sets were, of course, larger: there are only three ways to entangle a pair of qubits, and only one way to entangle a triple, for the 3-qubit system, while the number of ways for the 4-qubit system are 6 and 4, respectively, with one quad (and so on, following the binomial coefficients.) However the 4-qubit system required much less training: only 100 epochs at each stage. The 5-qubit system was trained for 100 epochs at each stage, but 98\% of the change in the parameters, as well as in the error reduction, had been accomplished by less than half that time, as shown in Figure~\ref{5q2err}. Successively larger systems still need to be trained to decrease initial errors; yet the amount of training necessary seems to diminish with successive training stages. 

\begin{figure}
\includegraphics[height=2in]{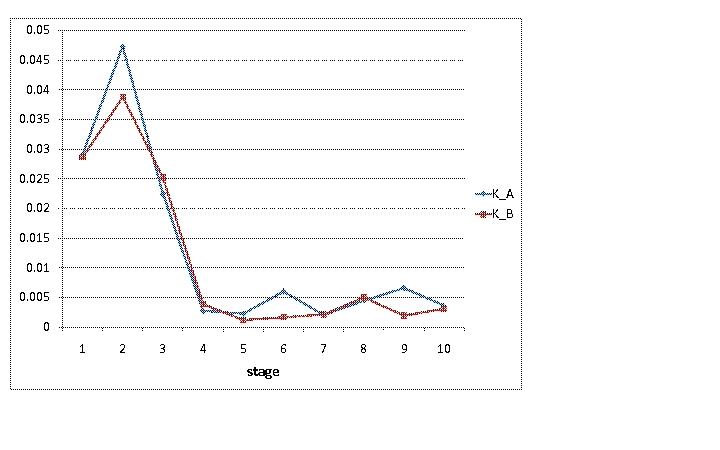}
\caption{Root square relative difference, $\sqrt{\sum{\Delta K^2}}/\sum{K}$, for both $K_{A}$ and $K_{B}$, as a function of training stage. The sums are over the timeslices, and the differences are to each successive stage. Stage 1 is 2-qubit training\cite{behrmanqic}; stage 2, 3-qubit pairwise; stage 3, 3q pairwise plus 3-way; stage 4, 4q pairwise; stage 5, 4q pairwise plus 3-way; stage 6, 4q pairwise plus 3-way plus 4-way; stage 7, 5q pairwise; stage 8, 5q pairwise plus 3-way; stage 9, 5q pairwise plus 3-way plus 4-way; stage 10, 5q pairwise plus 3-way plus 4-way plus 5-way. \label{Kev}}
 \end{figure}

\begin{figure}
\includegraphics[height=2in]{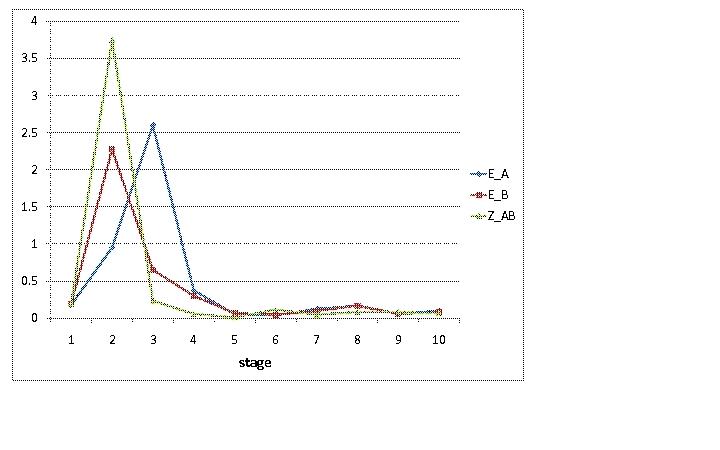}
\caption{Root square relative difference in each of the parameters $\varepsilon_{A}$, $\varepsilon_{B}$ and $\zeta_{AB}$ as functions of training stage. Stages as in Figure~\ref{Kev}. \label{ezev}}
\end{figure}

Figures~\ref{Kev} and \ref{ezev} make this point graphically. They show the evolution of the tunneling, bias, and coupling parameters, from initial training of the 2-qubit system, through each stage of the training of the 3-qubit, 4-qubit, and 5-qubit systems. Initially the QNN's learning entanglement requires large changes in the parameters, but as the size and complexity of the system is increased, the relative size of the changes decreases, till by stage 10 (the 5-qubit system's training to the full set of 56 training pairs), changes in the parameters are just a few percent. Note that in the two figures, the relative sizes of the final numbers (root square percent change in $K_{a}$ is 3.6\%; in $\zeta_{AB}$, 6.7\%) are comparable. This is reminiscent of the well-known neural network technique called ``bootstrapping''\cite{efron}, in which knowledge of part of a pattern allows tentative inference about the rest; in a sense, information about a smaller system can, in part, generalize to a larger. This makes us hopeful that this technique may be viable for use even on large systems.

\begin{figure}
\includegraphics[height=2.5in]{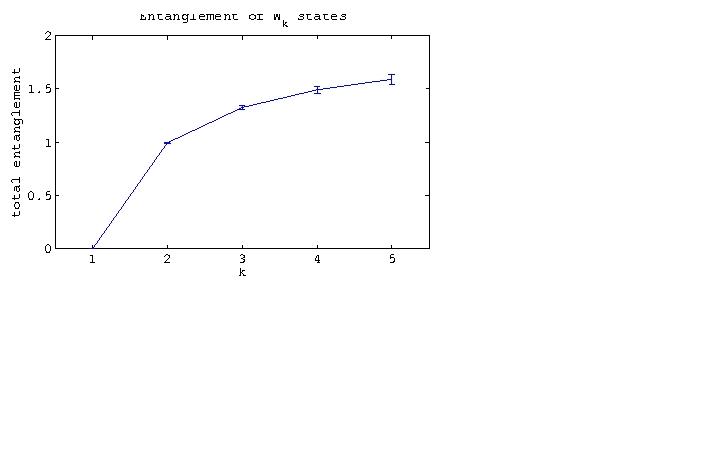} 
\caption{Total (pairwise) entanglement of W$_{k}$, calculated by the QNN using the trained parameters listed in Table~\ref{param}, as a function of k. Each data point is the average of the total entanglement for the $2^{k-1}\left(\begin{array}{c} 5 \\ k \end{array}\right)$ states of the form W$_{k}\otimes|0\rangle^{N-k}$. Error bars show the standard deviation at each point, which increases from from 0.7\%, for the $W_{2}$ state, to 2.8\% for the $W_{5}$ state. The line is drawn to guide the eye.\label{Wvsk}}
\end{figure}

Figure~\ref{Wvsk} shows the average total entanglement (sum over all outputs) of the W$_{k}$ states as a function of k, for k = 1 to 5, as computed by the QNN using the trained 5-qubit system. Since the amounts of 3-way, 4-way, and 5-way entanglements are zero on this scale, the total entanglement is equal to the sum of the pairwise entanglements for each state.  These numbers change only very slightly when we look at so-called ``flipped'' W states\cite{jung} instead (where, instead of a single excitation shared among N qubits, we have a single 0 so shared - e.g., $\tilde{W}_{3}= \frac{1}{\sqrt{3}}(|110\rangle\pm|101\rangle\pm|011\rangle)$). Recently Kimble et al.\cite{kimble} have succeeded in measuring the number of modes (k) sharing a single excitation photon; while the QNN is not, of course, calculating exactly this, our results do show similar separations among states of different k.

\begin{figure}
\includegraphics[height=2.5in]{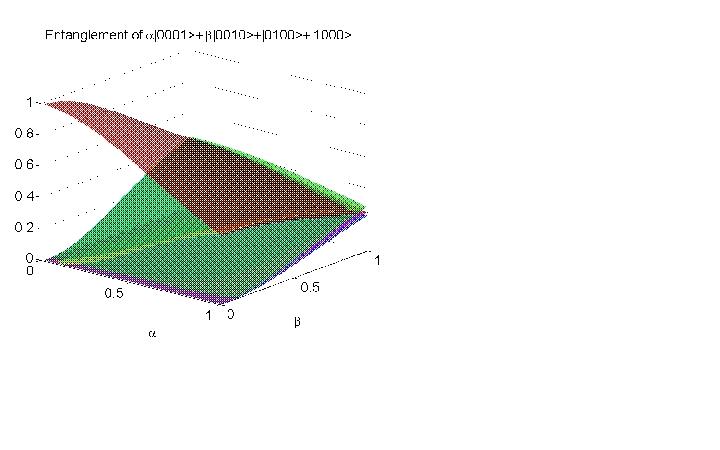} 
\caption{Pairwise entanglement of $\alpha|0001\rangle+\beta|0010\rangle+|0100\rangle+|1000\rangle$, as a function of $\alpha$ and $\beta$, as calculated by the QNN. Each color represents a different entanglement measure: red, the pairwise entanglement between A and B, $O_{AB}$; yellow, that between A and C, $O_{AC}$; green, $O_{AD}$; cyan, $O_{BC}$; blue, $O_{BD}$; and magenta, $O_{CD}$.\label{fourW}}
\end{figure}

Figure~\ref{fourW} shows the pairwise entanglements of the (normalized) states $\alpha|0001\rangle+\beta|0010\rangle+|0100\rangle+|1000\rangle$, calculated by the QNN using the trained 4-qubit system, as a function of both $\alpha$ and $\beta$. Results are not significantly different when using the trained 5-qubit system. When $\alpha=0$ and $\beta=0$, this is $(|01\rangle+|10\rangle)\otimes|00\rangle=EPR_{AB}\otimes|00\rangle$, which has full pairwise AB entanglement while all other entanglements are zero; our results match this. When $\alpha=1$ and $\beta=0$, this is $|0001\rangle+|0100\rangle+|1000\rangle$, which is the three-way W state in qubits ABD. In terms of pairwise entanglement, a three-way W state is equally pairwise entangled in all three possible combinations AB, AD, and BD, so we would expect the red, green, and blue surfaces to come together at that point, while the yellow (AC), cyan (BC) and magenta (CD) surfaces are zero there. This is in fact what we see in the figure. Similarly, when $\alpha=0$ and $\beta=1$, this is $|0010\rangle+|0100\rangle+|1000\rangle$, which is the three-way W state in qubits ABC; at this point the AB (red), AC (yellow), and BC (cyan) entanglements are all approximately equal and the AD (green), BD (blue), and CD (magenta) entanglements are zero. When both $\alpha=1$ and $\beta=1$, this is the four-way W state $|001\rangle+|0010\rangle+|0100\rangle+|1000\rangle$, equally entangled in all six pairwise ways AB (red), AC (yellow), AD (green), BC (cyan), BD (blue), and CD (magenta).

The value for the pairwise entanglement of the three-way W states deserves some further comment. In the original training set for the two-qubit system, it was found\cite{behrmanqic} that the value for the entanglement of the ``P'' state, $|01\rangle+|10\rangle+|11\rangle$, that resulted in the smallest overall error (training plus testing), was 0.44: that is, using this value, rather than some other, led to the greatest possible self-consistency for the QNN method. We therefore used this same value for $P\otimes|0\rangle$ for each of the three pairs, in training the three-qubit system\cite{behrmannabic}. The three-way W state $W_{ABC}=|001\rangle+|010\rangle+|100\rangle$ is not a ``P'' state in any of the three pairs, of course, but it is related in the following sense: we can write $W_{ABC}=|001\rangle+(|01\rangle+|10\rangle)\otimes|0\rangle$, that is, an EPR state in two of the qubits, plus a contamination term. In the same way, we can think of ``P'' as being a (fully entangled) EPR state, plus an amount of contamination, the inclusion of which diminishes the entanglement. Thus for self-consistency we would expect each of the pairwise entanglements of a W$_{3}$ state to be approximately 0.44, as in fact occurs.

\begin{figure}
\includegraphics[height=2.5in]{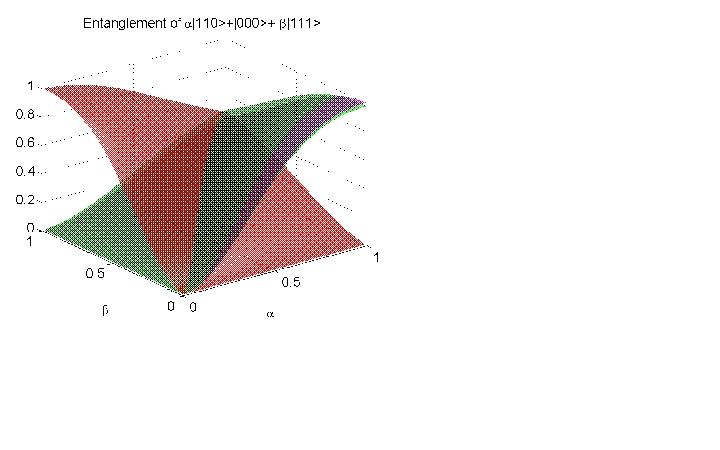} 
 \caption{Entanglement of $\alpha|110\rangle + \beta|111\rangle + |000\rangle$, as a function of $\alpha$ and $\beta$, as calculated by the QNN, and compared with the 3-tangle\cite{coffman}. The red surface shows the pairwise entanglement between qubits A and B, $O_{AB}$; the green, three-way entanglement among A, B, and C,$O_{ABC}$; and magenta is the 3-tangle.   \label{GHZBelltangle}}
 \end{figure}

\begin{figure}
\includegraphics[height=2.54in]{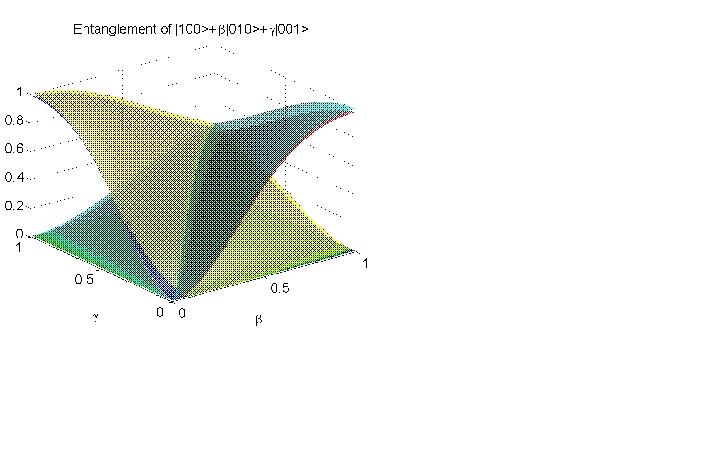} 
\caption{Entanglement of $|100\rangle + \beta|010\rangle + \gamma|001\rangle$, as a function of $\gamma$ and $\beta$, as calculated by the QNN, and compared with analytical results. Each color corresponds to a different entanglement measure: red, pairwise AB entanglement, $O_{AB}$; and blue, pairwise AC entanglement, $O_{AC}$, both computed by the QNN. Compare these to $\tau_{AB}$, the tangle between A and B, in cyan; and $\tau_{AC}$, the tangle between A and C, in yellow (both computed analytically.) The residual tangle is identically zero for all these states, in agreement with the QNN's calculated $O_{ABC}$, in green.\label{Wtangle}}
\end{figure}

We can also look at states which have three-way (or higher) entanglement. Figure~\ref{GHZBelltangle} shows the pairwise (red) and three-way (green) entanglement of the states $\alpha|110\rangle + \beta|111\rangle + |000\rangle$, calculated by the QNN using the trained 3-qubit system, as a function of $\alpha$ and $\beta$, and compared with the residual or 3-tangle of Coffman\cite{coffman} (magenta). Results were not significantly different when using the trained 4-qubit or five-qubit systems. When $\alpha = 0$ and $\beta = 1$, this is the pure GHZ state for three qubits, and the three-way entanglement is maximal; both the 3-tangle and the QNN calculate this to be one. Indeed the QNN three-way entanglement (green) tracks the 3-tangle (magenta) fairly well over the entire range of parameters shown. This can also be seen in Figure~\ref{Wtangle}, where we show the tracking of both pairwise and residual tangle by the QNN for states of the form $|100\rangle + \beta|010\rangle + \gamma|001\rangle$.

\begin{figure}
\includegraphics[height=2.5in]{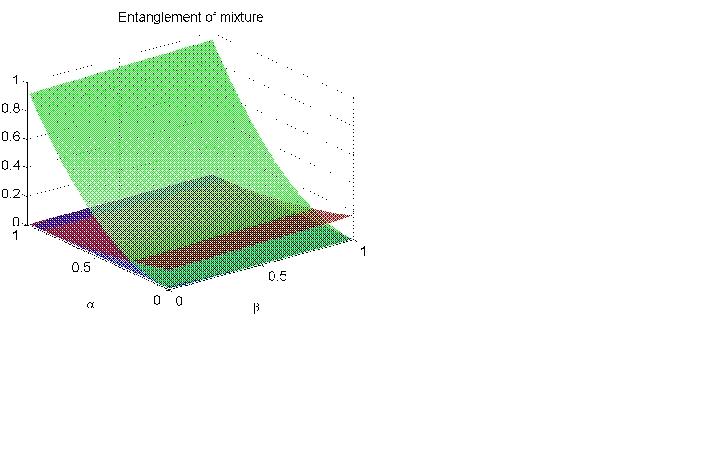} 
\caption{Entanglement of the mixed states $\alpha|GHZ_{5}\rangle\langle GHZ_{5}|+\beta|W_{5}\rangle\langle W_{5}|+(1-\alpha-\beta)|\tilde{W}_{5}\rangle\langle |\tilde{W}_{5}|$, as a function of both $\alpha$ and $\beta$.  Each color represents a different entanglement measure: red, the pairwise entanglement; blue, four-way entanglement; and green, five-way entanglement. \label{Wmx}}
\end{figure}

\begin{figure}
\includegraphics[height=2.5in]{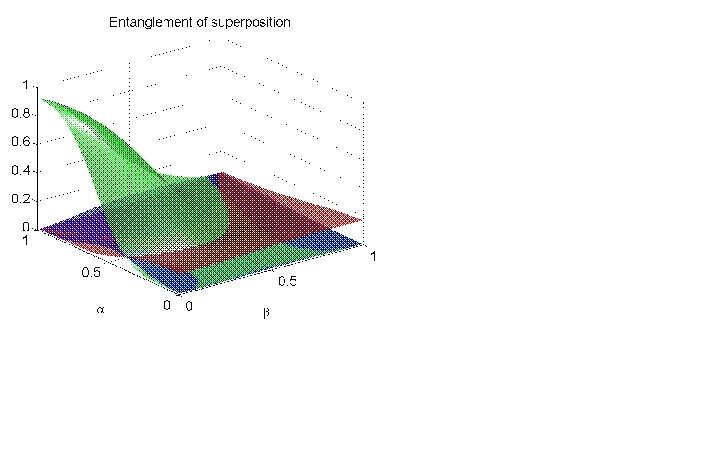} 
\caption{Entanglement of the superposition states $\alpha|GHZ_{5}\rangle+\beta|W_{5}\rangle+(1-\alpha-\beta)|\tilde{W}_{5}\rangle$, as a function of both $\alpha$ and $\beta$.  Each color represents a different entanglement measure: red, the pairwise entanglement; blue, four-way entanglement; and green, five-way entanglement. \label{Wsup}}
\end{figure}

With the trained net, we can evaluate any pairwise or N-wise entanglement for mixed states, as well, without having to find an optimal decomposition\cite{jung}. Figure~\ref{Wmx} shows the computed entanglement of states of the form $\alpha|GHZ_{5}\rangle\langle GHZ_{5}|+\beta|W_{5}\rangle\langle W_{5}|+(1-\alpha-\beta)|\tilde{W}_{5}\rangle\langle |\tilde{W}_{5}|$, as a function of both $\alpha$ and $\beta$.  As expected from symmetry, all the pairwise entanglements lie atop one another, so only one is shown, for clarity; all the three-way and four-way entanglements are zero. As this is a mixture, not a superposition, there is no interference among the three contributing states, and the $|W_{5}\rangle$ and $|\tilde{W}_{5}\rangle$ states contribute equivalently to the pairwise (red) entanglements.  Compare this relatively featureless behavior with that shown in Figure~\ref{Wsup}, which shows the calculated pairwise (red), four-way (blue), and five-way (green) entanglements of the superposition states $\alpha|GHZ_{5}\rangle+\beta|W_{5}\rangle+(1-\alpha-\beta)|\tilde{W}_{5}\rangle$. Again, all the pairwise entanglements (red) lie atop each other, by symmetry, as do the three-way (not shown) and the four-way (blue). The three-way entanglements are still all zero; however, the four-way (blue) are \underline{not}, because there are combinations of the coefficients such that the superposition contains some nonzero amplitude of each of the $|GHZ_{4}\rangle$ states; in neither $|W_{5}\rangle$ nor $|\tilde{W}_{5}\rangle$ are there any states with exactly three qubits in the same state, so there is no three-way entanglement. In both Figure~\ref{Wsup} and Figure~\ref{Wmx} we see the expected limiting behavior: for $\alpha$ = 1 and $\beta$ = 0, we have maximal five-way entanglement only; for $\alpha$ = 0 and $\beta$ = 1, we have maximal pairwise entanglement only, spread over all ten possible pairs (see also Figure~\ref{Wvsk}.)

\section{Conclusions}

We have presented a new approach for the estimation of pairwise and multiqubit entanglement, applicable to any state, whether pure or mixed. Using dynamic quantum backpropagation learning, we have shown that a quantum system can be trained to compute its own degree of entanglement. No prior state reconstruction or tedious optimization procedure is necessary, nor is ``closeness'' to any particular state. We envision an experimental implementation in which parameters could be, to begin with, roughly estimated from simulations such as those presented here (in the neural network literature, this is called ``offline'' or ``batch'' training\cite{wasserman}.) These parameters could then be refined experimentally (``online'' training of the neural network.) Because the state of the system at each timeslice must be known in order to do quantum backprop, it can only be done in simulation; some other parameter adjustment method would be needed experimentally, possibly a reinforcement or genetic algorithm. The experimental refinement procedure might also provide for the inclusion of features and interactions present in the physical state but not in the (simplified) model. In any case, once good values for the parameters are determined, any state's entanglement can be estimated by experimental measurement. And while we have only carried out simulations through a system size of 5 qubits, the additional training necessary significantly decreases as the size of the system grows, raising hopes for the method's feasibility even for large qubit networks.

It should be emphasized, though, that this method provides only an estimate of the entanglement (a ``witness'', not a ``measure.'') Agreement with analytical results, while good, is not exact. Also, we have not explored here application to phase shifted superpositions and mixtures, which present their own problems\cite{jung, lohmayer, bai, behrmanqic}, but which were not present in our training sets. Further work is needed in this area, and is ongoing.

That being said, we believe our present work is still of great value, providing as it does a way of bypassing difficult analytical work in addressing the need\cite{kimble} for experimental measurements to determine entanglement. Moreover, our approach may have other important advantages. Classically neural networks have proven fault tolerant and robust to noise; they are also famously used for noise reduction in signals. In quantum systems there is also the problem of decoherence. It is possible that quantum neural networks may be well suited for dealing with these types of problems in quantum computing. There are undoubtedly many other possible applications for the use of learning or AI methods as shortcuts to constructing quantum algorithms.

\begin{acknowledgments}
This work was supported in part by the National Science Foundation under Grant No. NSF PHY05-51164, through the KITP Scholars program (ECB), at the Kavli Institute for Theoretical Physics, University of California at Santa Barbara, Santa Barbara, CA. We thank W.K. Wootters, J.F. Behrman and J. Watkins for helpful discussions.
\end{acknowledgments}


\end{document}